# Sub-6 GHz Beam-Reconfigurable Microfluidic Antenna Using Graphene Liquid for 5G Network

Sasmita Dash[a], Constantinos Psomas[b], and Ioannis Krikidis[a]
[a]Department of Electrical and Computer Engineering, University of Cyprus, Nicosia, Cyprus
[b]Department of Computer Science and Engineering, European University, Nicosia, Cyprus

**Abstract**
As wireless communication systems continue to grow rapidly, high-performance antennas become increasingly crucial for expanding coverage, improving capacity, and enhancing transmission quality. In light of this, research has focused considerable attention on liquid antennas due to their unique characteristics, which include small size, flexibility, reconfigurability and transparency. Recently, graphene liquid has been explored for numerous applications due to its low cost, high conductivity, flexibility, and ease of processing. Specifically for antenna applications, graphene liquid performs better than conventional liquid metal. This paper presents a graphene-liquid antenna with beam reconfiguration ability for sub-6 GHz communication system. The graphene-liquid movement within the microfluidic channel is taken into consideration by the reconfiguration mechanism. The antenna achieves beam reconfiguration in 360° directions with 6 dBi of gain at 5.5 GHz, featuring a wideband impedance bandwidth of 24%. The antenna main beam is specifically reconfigured into six directions (0°, 45°, 135°, 180°, 225° and 315°) at 5.5 GHz. Additionally, in all six reconfigurable scenarios at 5.5 GHz, the antenna provides a stable reflection coefficient. Therefore, for the next generation of wireless communication systems, this novel design of graphene-liquid-based reconfigurable sub-6 GHz antennas holds promise.

## 1. Introduction

Wireless communication systems have experienced substantial revolutionary progress over the past few years. High data rates are the primary goal of deploying fifth-generation (5G) networks [1]. Nevertheless, the rapid growth of smart devices and heavy internet use will put a significant burden on 5G wireless networks. The sub-6 GHz bands and millimetre-wave (mmWave) spectrum have been used in the official commercialization of 5G. The sixth-generation (6G) technologies will further develop and expand the 5G paradigm in terms of data rate, coverage, connectivity, latency, positioning accuracy, and security [1]. The antenna plays a significant role in the present and next-generation communication systems. Significant technical developments in antenna design have occurred in tandem with the expansion of wireless communication networks to meet users' ever-increasing demands [2].

Conventional metal antennas offer remarkable performance but lack mechanical flexibility due to their typically rigid construction, which is often made of conductive metals on rigid substrates. Furthermore, these antennas lose their structural integrity and may even be destroyed if they are bent or stretched beyond a certain point [3].

Conversely, liquid metals in microfluidic channels preserve remarkable mechanical stability and flexibility without sacrificing their electrical characteristics [4]. Metallic liquid allows fluidic channels to take shape because of its low viscosity [5]. Their high degree of reversibility enables them to return to their initial state [4]. Liquid metals offer intrinsic flexibility, making them a potential alternative to solid conductors in flexible electronics [6]. Effective liquid antennas rely on the unique properties of liquid materials, which heavily influence their design and performance. Antennas that are significantly more flexible and reconfigurable are developed by using metallic liquids as radiative elements, rather than solid conductors, such as copper.

Graphene-liquid (GL), a new metallic liquid, has recently been used to develop metallic liquid antennas [7]. In the sub-6 GHz range, the radiation performance of the GL antenna outperforms that of conventional metallic liquid antennas in terms of gain, bandwidth, and radiation efficiency [7]. Currently, a wide range of industries use the liquid form of graphene [8]. Because of the significant research focus in recent years, graphene inks are being employed for Internet of Things, wireless connectivity, and flexible electronics applications [9, 10]. Graphene has established itself as a promising material for effective antenna design over the past decade [11, 12]. Graphene's stability and biocompatibility, along with its versatility, have made it a promising candidate for use in cardiac research, biomedical engineering, and neuroscience [13, 14]. However, the use of GL for antenna applications is underexplored. In this work, we present a beam-reconfigurable antenna operating at sub-6 GHz using GL inside a microfluidic channel.

## 2. Design and Analysis of Graphene-Liquid Antenna

In the sub-6 GHz frequency regime, we design and numerically analyze a GL antenna. Over platinum metal grounded liquid crystal polymer (LCP) ($\varepsilon_r$ = 2.9, tan $\delta$ = 0.0025) substrate (length Ls= 68 mm, width Ws=56 mm and height Hs=3 mm), the proposed antenna is realized by GL ($\approx$ 1 ml volume) in a rectangular-shaped polymethyl methacrylate (PM) ($\varepsilon_r$ = 2.55, tan $\delta$ = 0.002) microfluidic channel (length Lm = 46 mm, width Wm = 35 mm, and diameter Dm = 3 mm). The three-dimensional and side views of the proposed GL antenna for sub-6 GHz are shown in Fig. 1(a) and 1(b).

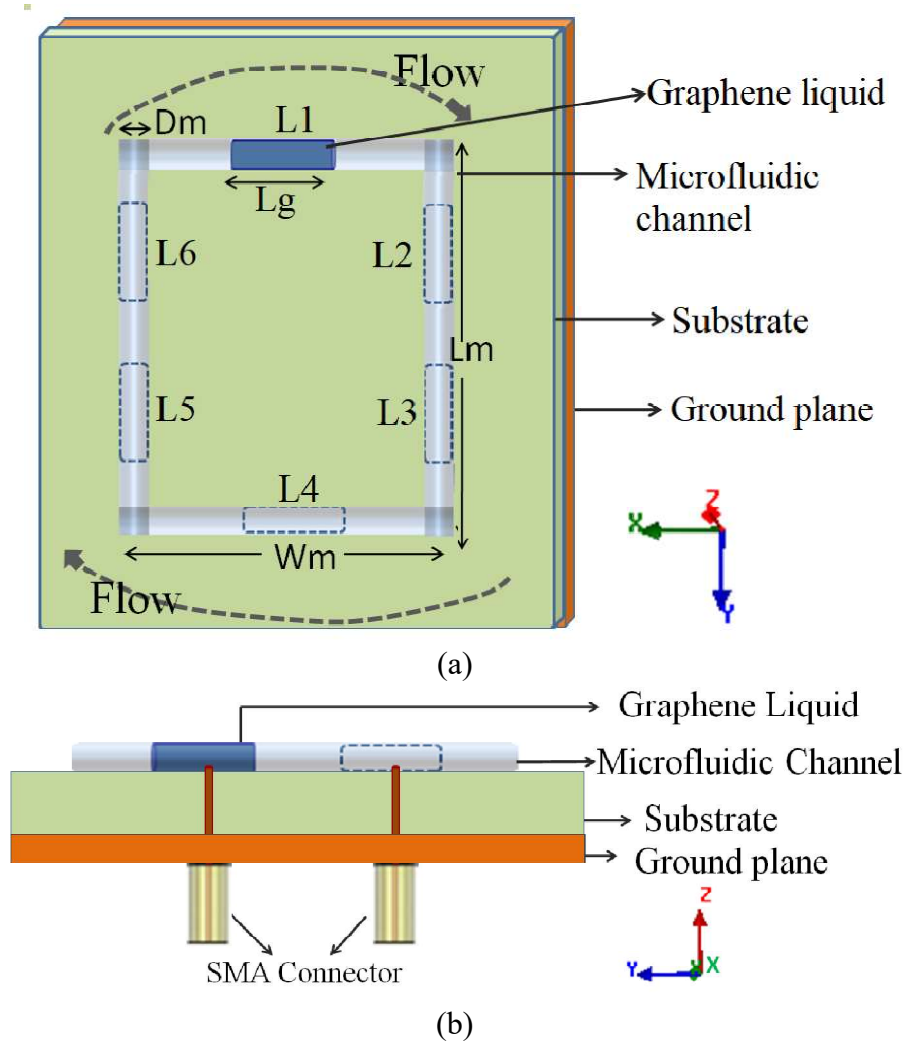

(a)

(b)

Figure 1: Schematic representation of the GL antenna. (a) 3D view and (b) side view.

Table 1 presents the dimensions of the proposed GL antenna for 5.5 GHz frequency. Ansys HFSS software is used to design, analyze and validate the performance of the proposed antenna [15]. The finite element method (FEM) based electromagnetic solver models GL as a conductive liquid at 5.5 GHz with intra-band conductivity using the Kubo formalism [16]. In the PM microfluidic channel, the GL moves from one location to another. Six locations of GL in the PM microfluidic channel are taken into consideration in this work. The antenna is excited using the centre-fed single-probe technique. Six feeding ports are made in six different locations to attain beam reconfigurability. It is possible to direct the antenna beam in six different directions by carefully choosing the location of the GL inside the PM microfluidic channel.

Table 1: The dimensions of the graphene liquid antenna at 5.5 GHz

| Ls (mm) | Ws (mm) | Lm (mm) | Wm (mm) | Dm (mm) | Lg (mm) |
|---------|---------|---------|---------|---------|---------|
| 68 | 56 | 46 | 35 | 3 | 12 |

Henceforth, the GL antenna reconfigure its radiation direction, covering 360° angle by movement of GL from one location to another inside the PM microfluidic channel. Whereas, in a Non-liquid Graphene antenna, beam reconfiguration and frequency reconfiguration are achieved through the electric field effect via a bias voltage, which tunes the surface conductivity of graphene [17, 18]. The following procedure can be used to fabricate the proposed microfluidic GL antenna. The GL antenna can be attained by injecting GL into a PM microfluidic channel over a metallic grounded LCP substrate [19, 20]. Microfluidic channels can be fabricated using soft lithographic techniques [21, 4, 20]. A thin LCP substrate layer can be used to seal the PM microfluidic channel of elastomer [20, 22].

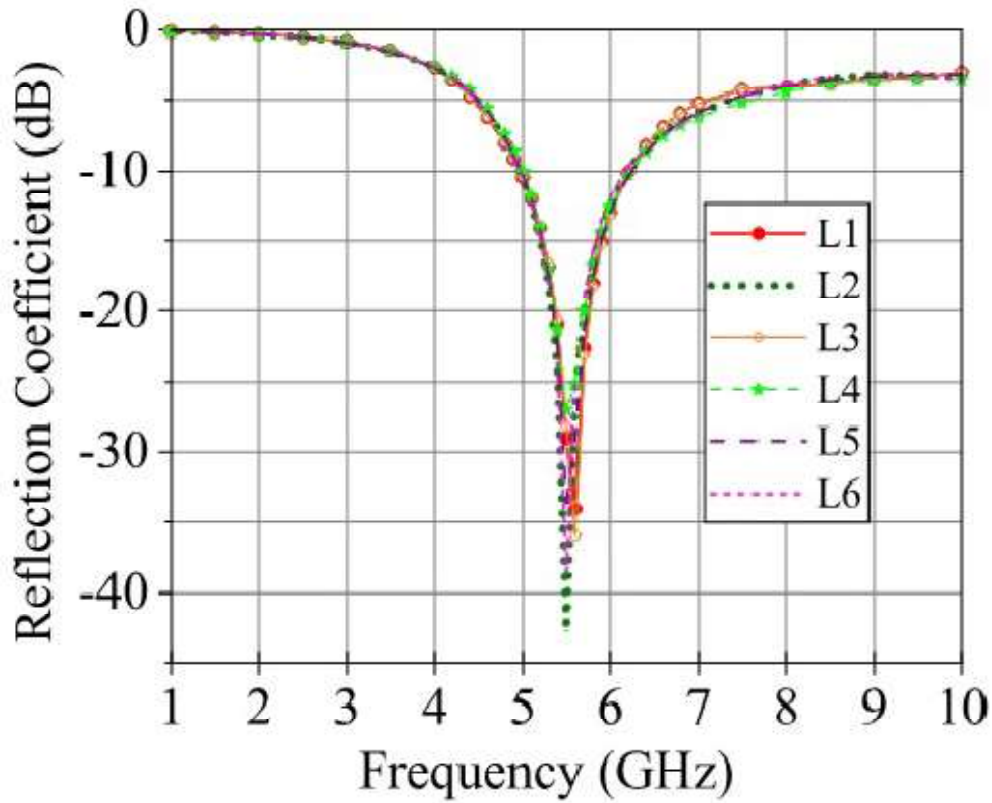

Figure 2: S11 parameter of the proposed antenna for six different positions of the graphene

The GL can first be injected into the PM liquid channel using a syringe. The antenna's liquid volume in the PM microfluidic channel will be located by the micropump unit. Microfluidic methods such as electrowetting or pumping can be used to displace the GL [22]. A silicon wafer sample can provide mechanical support for the GL antenna. Fig. 2 presents the S11 parameter of the GL antenna at six different locations in the microfluidic channel. From the Fig. 2, it can be seen that the antenna resonates at 5.5 GHz in six locations: L1, L2, L3, L4, L5, and L6. All six locations attain the same antenna resonant frequency and the bandwidth of the antenna is 24%. Six different operation states are made possible by the fluidic property of the GL in the PM microfluidic channel. The GL antenna achieves gain of 6 dBi and a symmetric unidirectional radiation pattern at 5.5 GHz, as illustrated in Fig. 3. Fig. 4 shows the normalized radiation pattern of the proposed GL antenna at 5.5 GHz. It is evident from Fig. 4 that the front-to-back ratio of the GL antenna is 10 dB, indicating lower backlobe radiation. The reconfiguration of the antenna beam is made possible by the GL moving to six different locations within the PM microfluidic channel, resulting in six different operational states. Fig. 4 shows the antenna radiation patterns in six operating modes. The antenna beam can be directed in B2 ($\theta = 0°$), B3 ($\theta = 45°$), B4 ($\theta = 135°$), B5 ($\theta = 180°$), B6 ($\theta = 225°$), and B1 ($\theta = 315°$) directions by carefully choosing the location of the GL inside the microfluidic channel. At an operating frequency of 5.5 GHz, the antenna can reconfigure its radiation direction to cover 360° angle.

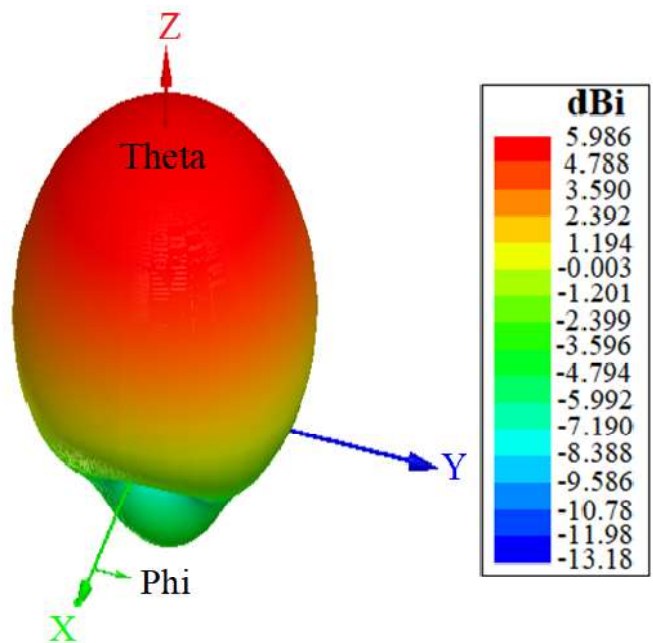

Figure 3: 3D far-field radiation pattern of the GL antenna at 5.5 GHz.

.

**Table 2: Main beam direction of the antenna in different position of Graphene-liquid.**

| Location of Graphene-liquid | L2 | L3 | L4 | L5 | L6 | L1 |
|---|---|---|---|---|---|---|
| Beam direction | B2 (0°) | B3 (45°) | B4 (135°) | B5 (180°) | B6 (225°) | B1 (315°) |

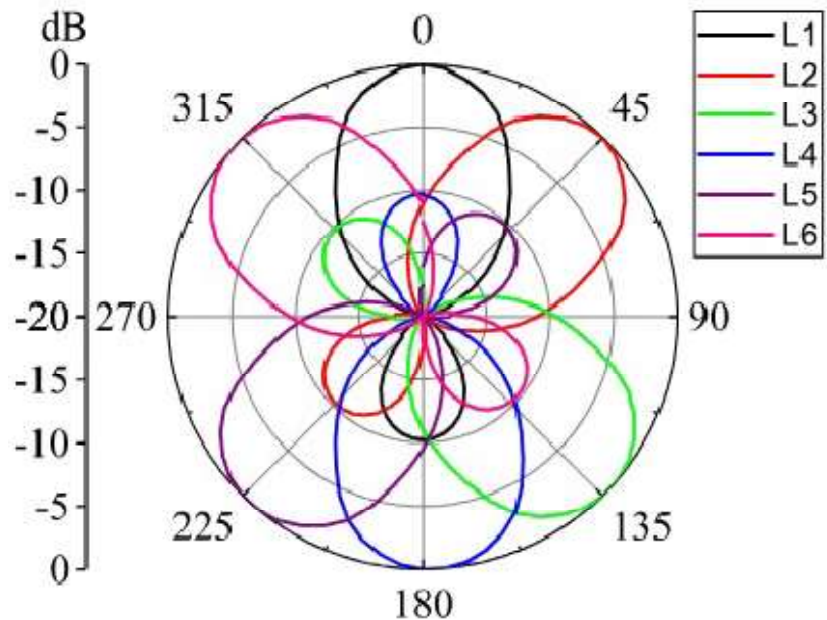

Figure 4: The normalized radiation pattern of the GL antenna in six operating states at 5.5 GHz.

The beam directions of the antenna with six locations of the GL are listed in Table 2. The normalized radiation patterns of the antenna in six working states are shown in Fig. 4. It is possible to direct the antenna beam in six different directions by carefully choosing the location of the GL inside the PM microfluidic channel. Henceforth, the GL antenna has the capability to reconfigure its radiation direction, covering 360° angle at 5.5 GHz.

## 3. Conclusion

This work presented a directional antenna capable of beam reconfiguration using GL in a PM microfluidic channel for sub-6 GHz wireless communications systems. The antenna is designed using GL in a PM microfluidic channel over LCP substrate. The mechanism of antenna reconfiguration is based on the movement of GL in the microfluidic channel. The antenna is beam reconfigured in 360° angle with six beams (0°, 45°, 135°, 180°, 225° and 315°) at 5.5 GHz. Additionally, the GL antenna achieves a gain of 6 dBi and a broad bandwidth of 24%. The results reveal that the microfluidically beam-reconfigurable antenna using GL shows promise for future use in wireless communications. The antenna will meet the demands of the constantly expanding network users, as next-generation wireless networks require high performance antennas.